\documentclass[twocolumn,english,aps,prl,superscriptaddress,showpacs,floatfix]{revtex4}
\usepackage{graphicx}
\usepackage{amsmath,amssymb,amsfonts}
\usepackage{natbib}
\usepackage{babel}
\usepackage{tikz}
\usetikzlibrary{arrows,shapes,positioning}

\tikzstyle{decision} = [diamond, draw, fill=blue!20, 
    text width=4.5em, text badly centered, node distance=3cm, inner sep=0pt]
\tikzstyle{block} = [rectangle, draw, fill=blue!20, 
    text width=7em, node distance=3cm, text centered, rounded corners, minimum height=4em]
\tikzstyle{line} = [draw, -latex']

\begin{document}

\title{Nonadiabatic \textit{ab initio} Quantum Dynamics without Potential Energy Surfaces}
\author{Guillermo Albareda} \email{guillermo.albareda@mpsd.mpg.de}
\address{Max Planck Institute for the Structure and Dynamics of Matter and Center for Free-Electron Laser  Science, Luruper Chaussee 149, 22761 Hamburg, Germany}
\address{Institute of Theoretical and Computational Chemistry, University of Barcelona, Mart\'i i Franqu\`es 1-11, 08028 Barcelona, Spain} 

\author{Aaron Kelly} \email{aaron.kelly@dal.ca}
\address{Max Planck Institute for the Structure and Dynamics of Matter and Center for Free-Electron Laser  Science, Luruper Chaussee 149, 22761 Hamburg, Germany} 
\address{Department of Chemistry, Dalhousie University, Halifax, Canada B3H 4R2}

\author{Angel Rubio}
\email{angel.rubio@mpsd.mpg.de}
\address{Max Planck Institute for the Structure and Dynamics of Matter and Center for Free-Electron Laser  Science, Luruper Chaussee 149, 22761 Hamburg, Germany} 
\address{Center for Computational Quantum Physics (CCQ), Flatiron Institute, 162 Fifth avenue, New York NY 10010}

\date{\today{}}

\begin{abstract}

We present an efficient \textit{ab initio} algorithm for quantum dynamics simulations of interacting systems that is based on the conditional decomposition of the many-body wavefunction [Phys. Rev. Lett. 113, 083003 (2014)]. Starting with this formally exact approach, we develop a stochastic wavefunction ansatz using a set of interacting conditional wavefunctions as a basis. 
We show that this technique achieves quantitative accuracy for a photo-excited proton-coupled electron transfer problem and for nonequilibrium dynamics in a cavity bound electron-photon system in the ultra-strong coupling regime, using two orders of magnitude fewer trajectories than the corresponding mean field calculation. This method is highly parallelizable, and constitutes a practical and efficient alternative to available quantum-classical simulation methods for systems of interacting fermions and bosons.   
\end{abstract}

\maketitle

New computational tools are still needed to treat nonequilibrium many-body quantum dynamics problems in molecular and condensed phase systems. Pragmatically speaking, while exact solutions are out of reach for a broad range of systems, our main goal is to provide a novel alternative theoretical framework for generating simulation-based predictions of observable properties that are as accurate as possible, in a computationally feasible, \textit{ab initio}, manner. 

Trajectory based quantum dynamics methods provide one possible route toward this goal, and offer the standard trade-off between physical accuracy and computational cost \cite{MillerPerspective,TullyPerspective,kapralreview}. Of these approaches, perhaps the most popular are Ehrenfest mean field theory\cite{maclachlan}, and Tully's surface hopping dynamics\cite{tully90}. Both of these approaches can be simulated using an ensemble of uncorrelated trajectories. Reintroducing correlation, for example by using wavepacket methods \cite{FMS,g-MCTDH,vmcg,MCE,cloning}, semiclassical techniques\cite{scivr,fbivr}, the quantum-classical Liouville equation \cite{qcle,mqcle,jfbts}, linearization-based approaches\cite{ShiGeva-linear,landmap,ildm,pldm}, or methods based on the exact factorization \cite{exact_fac,ct-mqc,dish-xf}, allows for further accuracy at the expense of (often significant) additional computational effort. In practice, essentially all of these quantum dynamics methods are formulated using a discrete (truncated) Hilbert space representation for the electronic degrees of freedom. In this picture, the Born-Oppenheimer (BO) approximation naturally emerges as classical nuclear dynamics on the electronic ground state potential energy surface (BOPES) \cite{BO}, and nonadiabatic effects are introduced by including multiple electronic potential energy surfaces, and nonadiabatic coupling terms (NACTs) \cite{domcke}. 

An alternative to this approach, is to use the (real space) position representation for the electrons. This allows one to go beyond the BO picture, without the need to explicitly calculate several BOPESs and NACTs\cite{PRA}. This is an attractive feature from a computational point of view, as these quantities may be demanding to obtain from \textit{ab initio} electronic structure calculations. The conditional wavefunction (CWF) approach can be formulated in this picture; it is an exact decomposition and recasting of the unitary time-evolution of a closed quantum system, that yields a set of coupled, non-Hermitian, equations of motion \cite{PRL1}. Inspired by the trajectory based approach to quantum dynamics of de Broglie and Bohm \cite{Bohm,gindensperger,Ivano,Sophya}, the CWF approach allows one to describe the evolution of arbitrary subsets of the degrees of freedom in a system, on a formally exact level. In addition, this alternative formulation of the many-body quantum dynamics problem allows novel approximate schemes to be developed \cite{JPCL1,Tarek} providing a completely new perspective to deal with the long-standing problems of nonadiabatic dynamics in complex interacting systems. 

In this Letter we report an approach for performing nonadiabatic quantum dynamics simulations using a set of time-evolving basis functions that are obtained from an approximation to the exact CWF equations of motion. This technique allows one to bypass the, typically necessary, computation of multiple BOPESs and NACTs, and potential subsequent diabatization procedures.  Hence, this method offers a new and attractive route to calculate observables and time correlation functions without relying on the widely used concept of the BOPES.

We consider a closed system of interacting particles, and separate the degrees of freedom into two arbitrary subsets. We also use the position representation for both sets; lowercase symbols will be used for the first subsystem, e.g. $\mathbf{r} = \{ \mathbf{r}_1,...,\mathbf{r}_{n} \}$, and uppercase symbols $\mathbf{R} = \{ \mathbf{R}_1,...,\mathbf{R}_{N} \}$ for the second. This decomposition holds for an arbitrary number of subsets (up to the total number of degrees of freedom in the system), and applies to both fermionic and bosonic many-body interacting systems. Here we choose $n$ and $N$ to represent the total number of degrees of freedom in each of the two subsystems. 

The conditional wavefunction (CWF) approach can be developed starting from the full time-dependent Schr\"odinger equation (TDSE) in the position representation,\begin{equation}\label{TDSE}
	i \hbar \frac{\partial}{\partial t} \Psi(\mathbf{r}, \mathbf{R}, t) = \hat{H}(\mathbf{r},\mathbf{R},t)\Psi(\mathbf{r}, \mathbf{R}, t).
\end{equation}
The total Hamiltonian for the system is\begin{equation}
\hat{H} =\hat{T}_1(\mathbf{r}) + \hat{T}_2(\mathbf{R}) + W(\mathbf{r},\mathbf{R},t),
\end{equation}where the kinetic energy operators for each species $j$ are $\hat{T}_j=\frac{\hbar^2}{2m_j}(i\nabla_j - \textbf{A}_j(t))^2$, and $m_j$ are their characteristic masses. The vector potential (in Coulomb gauge) due to an arbitrary external electromagnetic field, $\textbf{A}_j(\mathbf{r}, \mathbf{R}, t)$ is also included. The full interacting potential energy of the system is $W(\mathbf{r},\mathbf{R},t)$. 

The total wavefunction can be exactly decomposed in terms of the CWFs of either of the two subsystems,\begin{eqnarray}\label{CWF1}
	\psi_{1}^\alpha(\mathbf{r},t) &:=&
 \int d\mathbf{R} \delta(\mathbf{R}^\alpha(t) - \mathbf{R}) \Psi(\mathbf{r},\mathbf{R},t),	
\\ \label{CWF2}
	\psi_{2}^\alpha(\mathbf{R},t) &:=& 
 \int d\mathbf{r} \delta(\mathbf{r}^\alpha(t) - \mathbf{r}) \Psi(\mathbf{r},\mathbf{R},t).
\end{eqnarray}
Using these definitions in Eq. (\ref{TDSE}), one can show that the CWFs, $\psi_{1}^\alpha(t)$ and $\psi_{2}^\alpha(t)$, obey the following equations of motion:\begin{eqnarray}\label{conditional_e}
 i\hbar \frac{d \psi_{1}^\alpha}{d t} &=& \left(  \hat{T}_1 (\mathbf{r}) + W(\mathbf{r},\mathbf{R}^\alpha,t) +
 \eta_1^\alpha(\mathbf{r},t) \right) \psi_1^\alpha,
\\ \label{conditional_n}
 i\hbar \frac{d \psi_{2}^\alpha}{dt} &=& \left(  \hat{T}_2(\mathbf{R}) + W(\mathbf{r}^\alpha,\mathbf{R},t) + \eta_2^\alpha(\mathbf{R},t) \right) \psi_2^\alpha,
\end{eqnarray}where we have suppressed the explicit time-dependence of the coordinates, 
i.e., $\left\{\mathbf{r}^{\alpha},\mathbf{R}^{\alpha}\right\} \equiv \left\{\mathbf{r}^{\alpha}(t),\mathbf{R}^{\alpha}(t)\right\}$. The complex potentials $\eta_1^\alpha(\mathbf{r},t) $ and $\eta_2^\alpha(\mathbf{R},t) $ are functionals of the full wavefunction, and are given in Refs.\cite{PRL1,PRA}.
The conditional wavefunctions, (\ref{CWF1}) and (\ref{CWF2}), represent \textit{slices} of the full wavefunction taken along the degrees of freedom of the two disjoint subsets (see, e.g., Fig. 2 in Ref. \cite{PRA}). Each individual conditional wave function constitutes an open quantum system, whose time-evolution is non-unitary, due to the complex potentials.

An exact solution to Eq. (1) can be constructed provided an ensemble of trajectories $\{ \mathbf{r}^\alpha,\mathbf{R}^\alpha \}$ that explores the full support of $|\Psi(\mathbf{r},\mathbf{R},t)|^2$. For example, an ensemble of Bohmian trajectories defined through the conditional velocity fields \cite{EPJD,PRLoriols} would fulfill such requirements.
An approximate solution can be formulated \cite{PRL1} by expanding the complex functionals around the conditional coordinates, and then truncating such that $\eta_1^\alpha (\mathbf{r},t) = f(\mathbf{R}^\alpha,t)$ and $\eta_2^\alpha (\mathbf{R},t) = g(\mathbf{r}^\alpha,t)$. In this limit, these potentials only engender a pure time-dependent phase that can be omitted, as the conditional velocity fields are invariant under such global phase transformations \cite{PRL1}. The resulting propagation scheme is thus restored to a Hermitian form, and this approximate version of the CWF formalism is referred to as the Hermitian-CWF approach \cite{PRL1}. 

The Hermitian-CWF propagation scheme recasts the full quantum time-propagator as a set of independent single-species propagators, which is clearly a major simplification of the full problem. Hence, this form of the conditional decomposition allows one to circumvent the problem of storing and propagating the full many-body wavefunction, whose size scales exponentially with the number of degrees of freedom. 

In this Letter we consider the following ansatz for the full many-body wavefunction: \begin{equation}\label{ansatz}
	\Psi(\mathbf{r},\mathbf{R},t) = \sum_{\alpha=1}^{M} C_\alpha(t) \psi_1^\alpha(\mathbf{r},t)
    \psi_2^\alpha(\mathbf{R},t).
\end{equation} The basis functions in this sum are chosen to be Hermitian-CWFs, and the upper limit of the sum, $M$, refers to the total number of stochastically sampled trajectories (that we will show below can be kept to a very low number, making the present scheme computationally very efficient). Including interactions between the trajectories in the ensemble corrects the Hermitian-CWF evolution, through the set of complex time-dependent coefficients, $\mathbf{C}(t) = \left\{ C_1(t),...,C_M(t) \right\}$. The time evolution of these coefficients, is obtained by inserting Eq. (\ref{ansatz}) into Eq. (\ref{TDSE}), \begin{equation}\label{coeff}
	i\mathbb{M} \dot{\mathbf{C}}(t) = \left(\mathbb{W} - \mathbb{W}_1 - \mathbb{W}_2 \right)\mathbf{C}(t),
\end{equation} where the matrix elements of $\mathbb{M}$, $\mathbb{W}$, $\mathbb{W}_1$, and $\mathbb{W}_2$ are:
\begin{subequations}\label{Matrices}
\begin{align}
 \label{M}
 \mathrm{M}_{\alpha,\alpha'} &= \int d\mathbf{r} \psi_1^{\alpha '*} \psi_1^\alpha
    \int d\mathbf{R} \psi_2^{\alpha '*} \psi_2^\alpha \\
 \label{W}   
 \mathrm{W}_{\alpha,\alpha'} &= \int d\mathbf{r}d\mathbf{R} \psi_1^{\alpha '*} 	
    \psi_1^\alpha \psi_2^{\alpha '*} \psi_2^\alpha W(\mathbf{r},\mathbf{R})\\
 \label{W1}   
 \mathrm{W}_1^{\alpha,\alpha'} &= \int d\mathbf{r} \psi_1^{\alpha '*} \psi_1^\alpha 	
    W(\mathbf{r},\mathbf{R^\alpha}) \int d\mathbf{R} \psi_2^{\alpha '*} \psi_2^\alpha, \\
 \label{W2}   
 \mathrm{W}_2^{\alpha,\alpha'} &= \int d\mathbf{r} \psi_1^{\alpha '*} \psi_1^\alpha 	
    \int d\mathbf{R} \psi_2^{\alpha '*} \psi_2^\alpha W(\mathbf{r^\alpha},\mathbf{R}).
\end{align}
\end{subequations} 

Obtaining these matrix elements is straightforward and, except for \eqref{W}, they can be easily calculated from independent single species integrals. Evaluating the matrix elements of $\mathbb{W}$, in principle, requires the reconstruction of the full (ansatz) wavefunction. This does not restrict the use of the method to cases where the potential energy $W(\mathbf{r},\mathbf{R})$ can be fit to a sum-of-products form, as in the multi-configurational time-dependent Hartree method \cite{MCTDH_review} for example, but it does pose a potential numerical challenge in the case of a large trajectory ensemble. 

Once the coefficients $\mathbf{C}(t)$ are known, the velocity fields $\{\dot{\mathbf{r}}^\alpha, \dot{\mathbf{R}}^\alpha \}$ are then constructed according to the exact expressions for each subsystem:
\begin{eqnarray}\label{v1}
     \dot{\mathbf{r}}^\alpha(t) = \text{Im}\left[  \frac{\sum_\alpha C_\alpha(t) \psi_2^\alpha(\mathbf{R}^\alpha,t) \left( \nabla_\mathbf{r} \psi_1^\alpha(\mathbf{r},t) \right)|_{\mathbf{r}^\alpha(t)}}{\sum_\alpha C_\alpha(t) \psi_1^\alpha(\mathbf{r}^\alpha,t)\psi_2^\alpha(\mathbf{R}^\alpha,t)}  \right], 
\\ \label{v2}
     \dot{\mathbf{R}}^\alpha(t) = \text{Im}\left[  \frac{\sum_\alpha C_\alpha(t) \psi_1^\alpha(\mathbf{r}^\alpha,t) \left( \nabla_\mathbf{R} \psi_2^\alpha(\mathbf{R},t) \right)|_{\mathbf{R}^\alpha(t)}}{\sum_\alpha C_\alpha(t) \psi_1^\alpha(\mathbf{r}^\alpha,t)\psi_2^\alpha(\mathbf{R}^\alpha,t)}  \right]. 
\end{eqnarray}

The Interacting-CWF method, described above, does not require the electronic BOPES or NACs as input, or for time propagation. This feature is potentially quite advantageous for treating processes that involve many quantum states or continua, as in light-induced dynamics or surface-scattering phenomena. In addition, the Interacting-CWF propagation scheme avoids the computation of the nonlocal complex potentials, $\eta_1^\alpha(\mathbf{r},t) $ and $\eta_2^\alpha(\mathbf{R},t) $, as it is based on the Hermitian limit of the CWF equations of motion. Furthermore there is minimal cross-talk between trajectories, which makes the algorithm computationally efficient in massively parallel architectures \cite{SIA}.

\begin{figure}
  \includegraphics[width=0.5\textwidth]{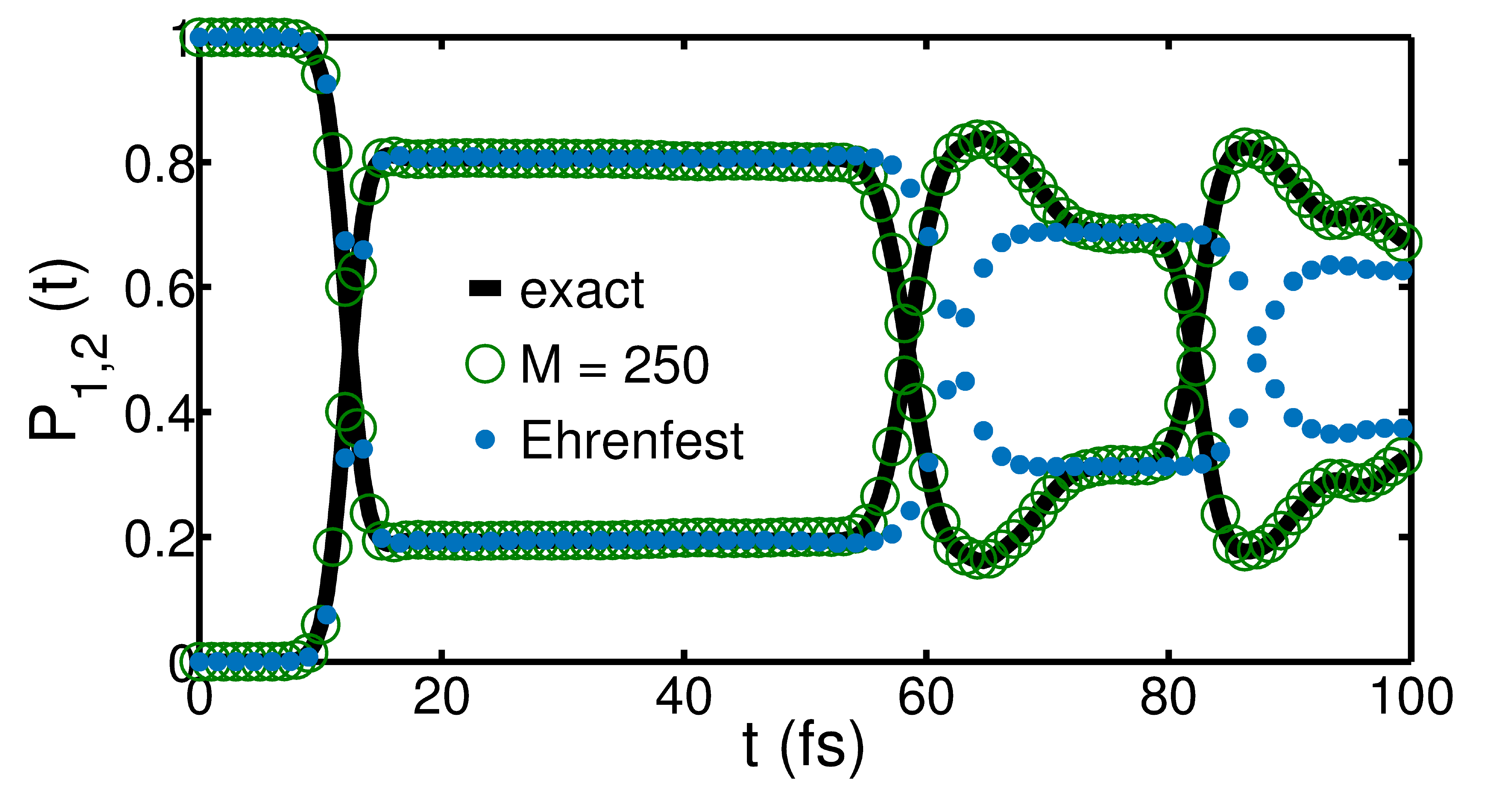} 
  \includegraphics[width=0.5\textwidth]{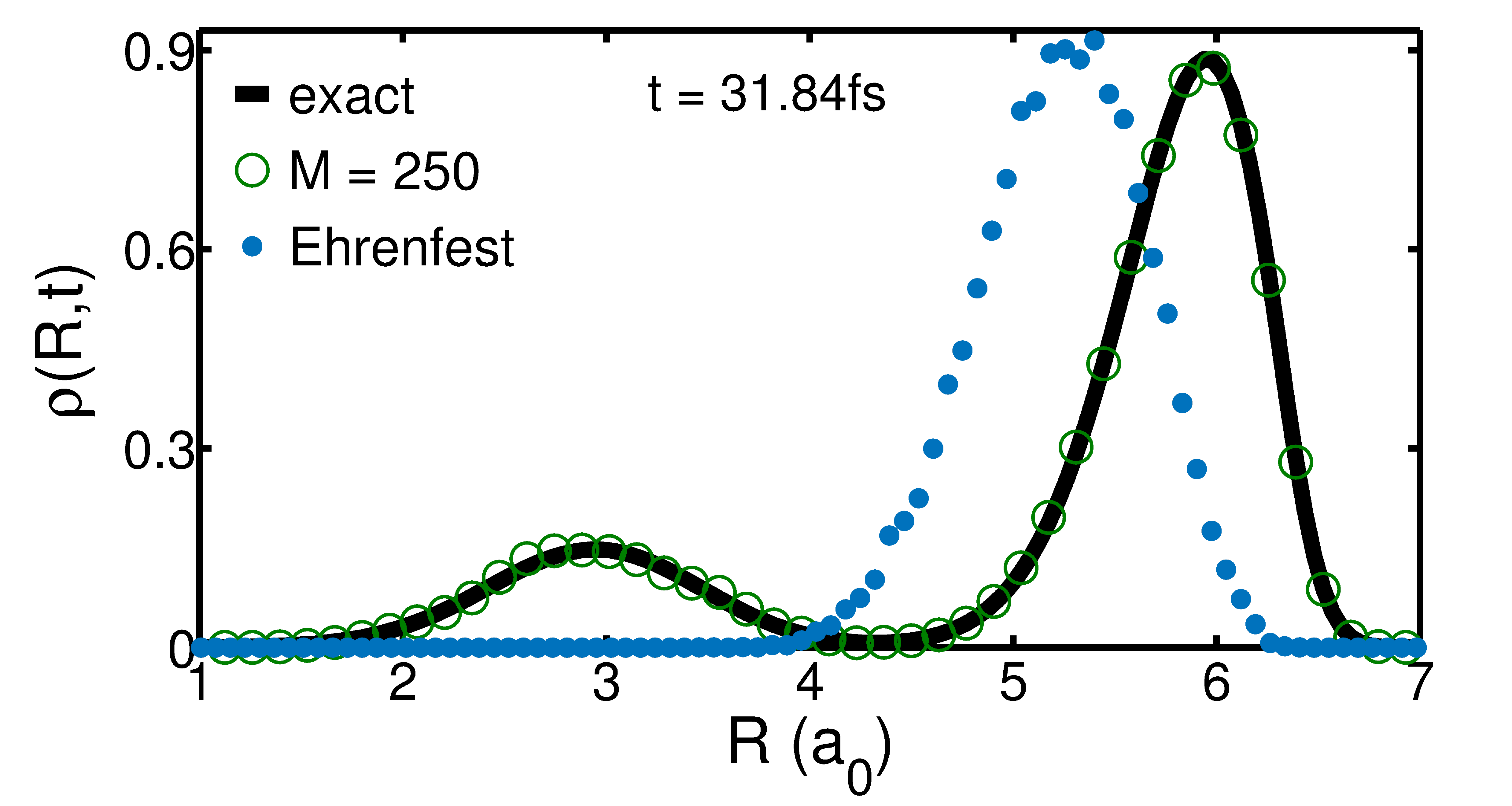}
  \includegraphics[width=0.5\textwidth]{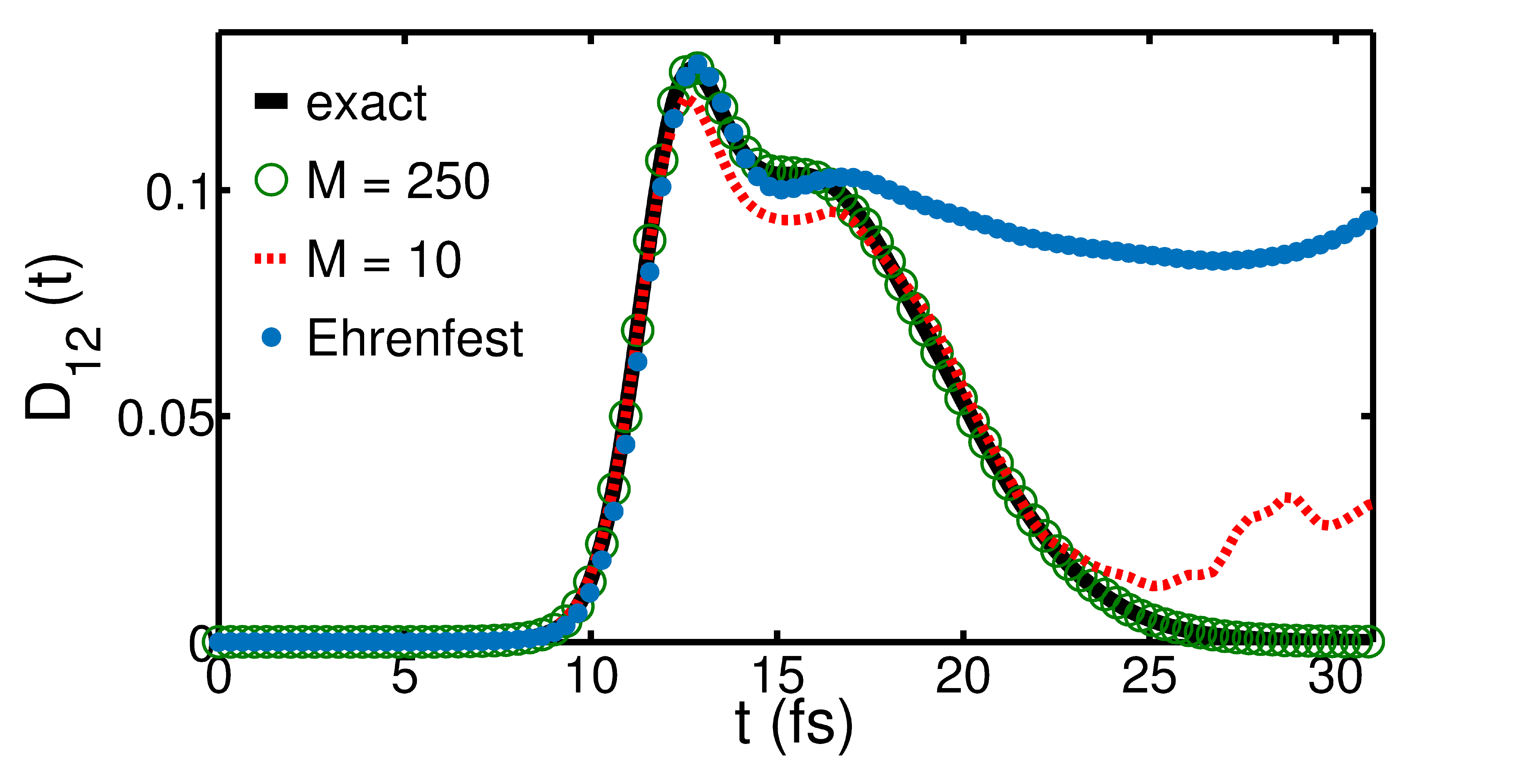}
  \caption{Dynamics in the Shin-Metiu model for photoinduced proton-coupled electron transfer (see text and SI for details). Exact results (black line), Interacting-CWF with $M=250$ (open green circles) and Ehrenfest mean field theory (blue dots). Top panel: Evolution of the BO state populations $P_1(t)$ and $P_2(t)$. Mid panel: Snapshot of the reduced nuclear coordinate density $\rho_n(R,t)$ at $t = 31.84fs$. Bottom panel: Time-dependent decoherence indicator $D_{12}(t)$; Interacting-CWF results with 10 trajectories (red dashed line).}
\label{shin-metiu}
\end{figure}

We first show Interacting-CWF simulation results for a prototypical photo-induced proton-coupled electron transfer reaction, using the Shin-Metiu model \cite{Metiu}. The system comprises donor and acceptor ions which are fixed in space, and a proton and an electron that are free to move in one dimension along the line connecting the donor-acceptor complex. This model is very flexible and describes situations where electron-nuclear correlations play a crucial role in the reaction dynamics \cite{SIB}. The parameters of the model are chosen to make contact with previous work on non-adiabatic relaxation for this system \cite{PRL1,Ali2}.   

We choose an initial state for the system with the electron in the first excited BO state, while the initial nuclear wavefunction is chosen to be a Gaussian centered on the equilibrium geometry of the ground state. The short-time dynamics that proceeds from this initial condition involves a passage through an avoided crossing of two BOPESs, with further crossings occurring at later times as the system evolves. In order to characterize the dynamics, we monitor the BO electronic state populations, the reduced nuclear probability density, $\rho(\mathbf{R},t) = \int d\mathbf{r} |\Psi(\mathbf{r},\mathbf{R},t)|^2$ as well as an indicator of decoherence that is defined as the overlap integral of projected nuclear densities evolving on different BO electronic states, $D_{nm}(t) = \int d\mathbf{R} |\chi^{(n)}(\mathbf{R},t)|^2|\chi^{(m)}(\mathbf{R},t)|^2$. 

When the system passes through the nonadiabatic coupling region, the electron transfers probability between the first excited state and the ground state (top panel of Fig. \ref{shin-metiu}). As a result of the electronic transitions, the reduced nuclear density changes shape by splitting into two parts representing influences from both ground and excited state BOPES's at $\text{t}\approx 32$fs (middle panel of Fig. \ref{shin-metiu}). As non-adiabatic transitions occur the system builds up a degree of coherence, and this coherence subsequently decays as the system evolves away from the coupling region (bottom panel of Fig. \ref{shin-metiu}). The Interacting-CWF method vastly outperforms the Hermitian-CWF approach \cite{PRL1}, Ehrenfest mean field theory (also shown in Fig. \ref{shin-metiu}), as well as Tully's surface hopping dynamics \cite{fede}, in describing all these aspects of this problem. While both the Interacting-CWF method and Ehrenfest dynamics correctly capture the exact population dynamics at short times, the latter breaks down at longer times. Mean field theory also fails to capture the qualitative structure of the time-evolving reduced nuclear density, and the indicator of decoherence. These features are perfectly captured by our Interacting-CWF approach using very few trajectories; fully converged Interacting-CWF results were reached with $250$ trajectories, while $2$x$10^4$ trajectories (initially sampled from the Wigner distribution corresponding to the initial quantum nuclear wave packet) were required for convergence with Ehrenfest dynamics. Somewhat surprisingly, as shown in the bottom panel Fig. 1, electronic decoherence is captured nearly quantitatively by the Interacting-CWF method using only 10 trajectories.

Next, we simulate a single electron in a one-dimensional double well potential that is coupled to a quantum electrodynamical (QED) cavity through a single photon mode in the ultra-strong coupling regime \cite{PNAS1,PNAS2,Nature}. This scenario constitutes a formidable challenge for approximate approaches. In this regime the ground BOPES is strongly coupled to the first excited state, and there are significant couplings among the other higher-lying excited states \cite{SIC}. The effective mass of the photon displacement coordinate is identical to the electronic mass, hence the dynamics deviate strongly from the BO limit. Furthermore, tunneling, quantum coherence, and zero-point energy conservation are also important for both interacting subsystems. 
\begin{figure}
  \includegraphics[width=0.5\textwidth]{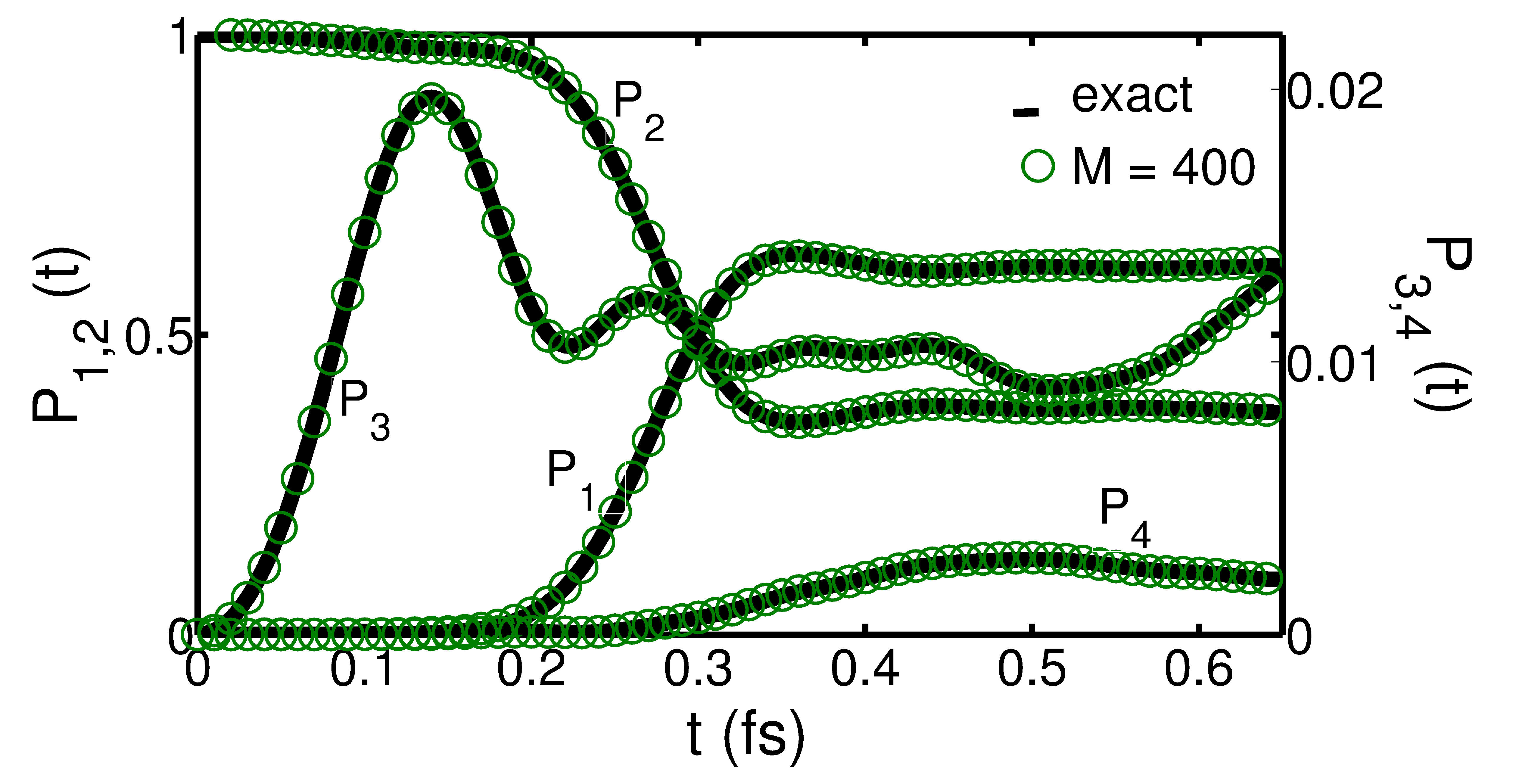}
  \includegraphics[width=0.5\textwidth]{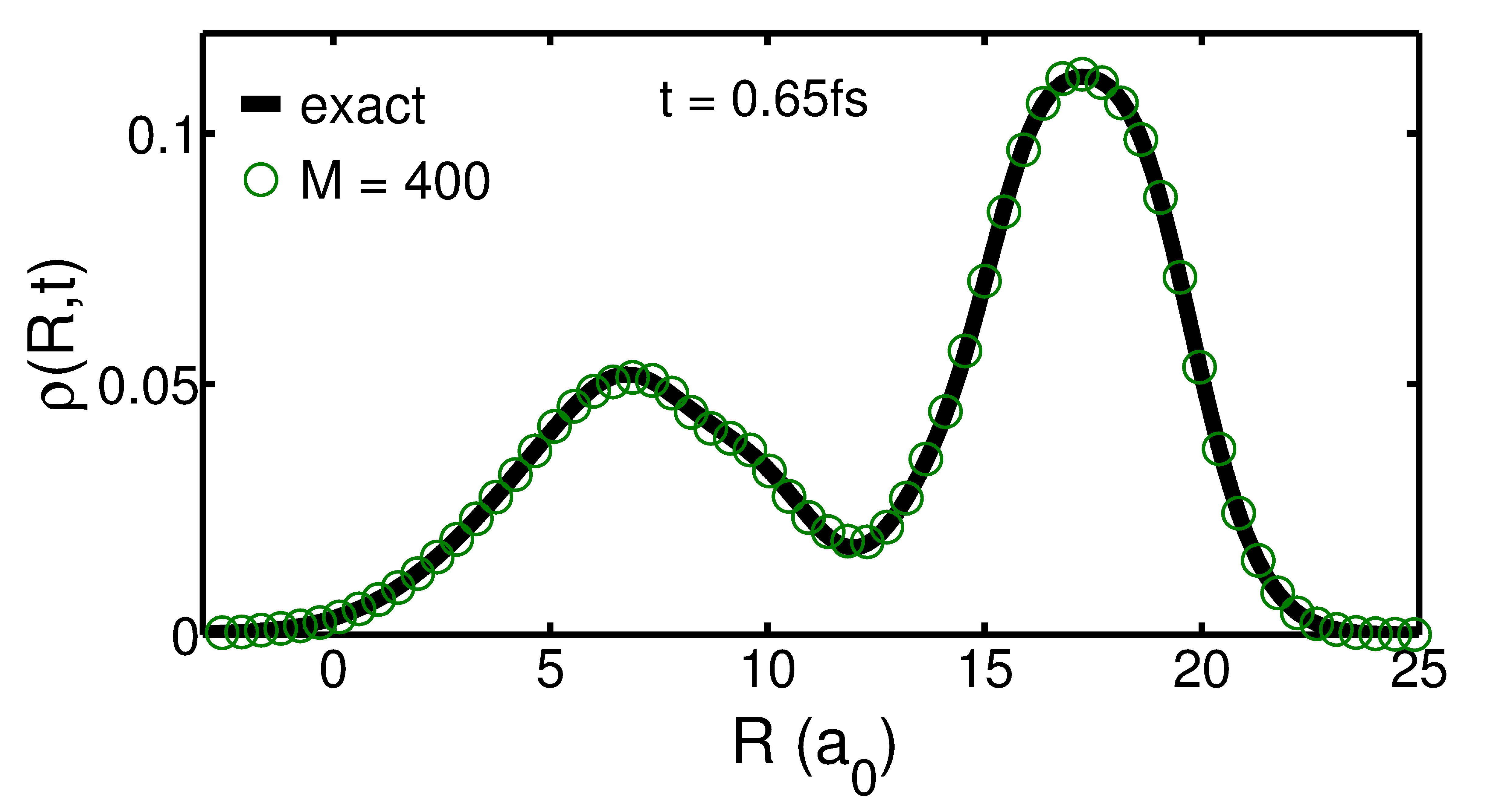}
  \includegraphics[width=0.5\textwidth]{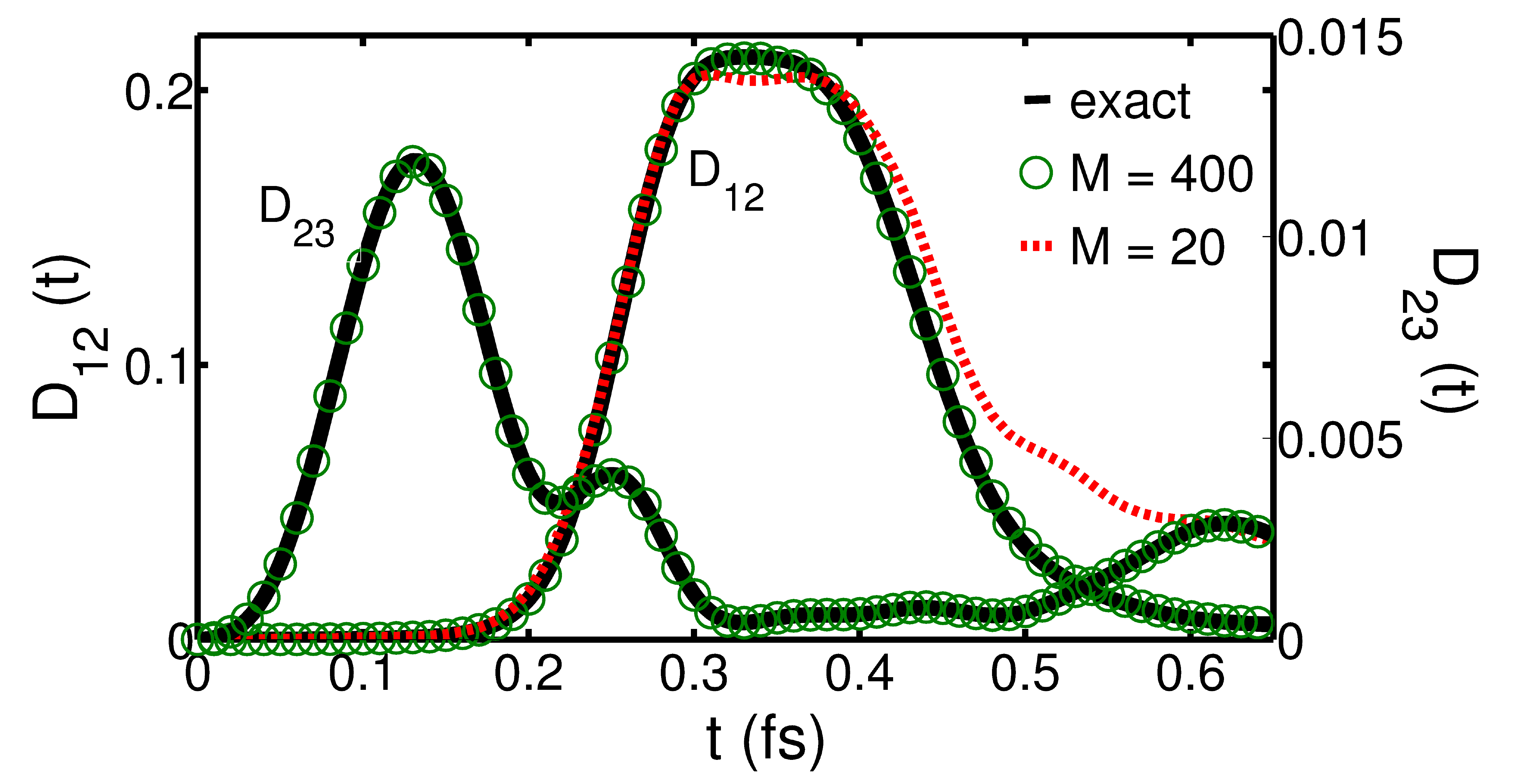}
  \caption{Nonequilibrium QED cavity-bound electronic dynamics in the ultra-strong coupling regime. Exact (black solid line) and Interacting-CWF (green circles) with 400 trajectories. Top panel: Evolution of electronic BO state populations. Mid panel: Reduced quantized displacement-field coordinate probability density $\rho(R,t)$ at $t = 0.65fs$. Bottom panel: Time-dependent decoherence indicator $D_{12}(t)$ and $D_{23}(t)$; Interacting-CWF results with 20 trajectories (red dashed line).}
  \label{el_ph}
\end{figure}

We choose the system to be initially prepared with the electron in the first excited BO state, while the initial photonic wavefunction is prepared in a displaced coherent state (see \cite{SIC}). This initial state is not a stationary eigenstate of the full electron-photon Hamiltonian, and it evolves in time by scattering thorough multiple avoided crossings. The reduced photon density develops a complex structure as time progresses as there are at least four BOPESs involved in the dynamics (see top panel of Fig. (\ref{el_ph})). The exact electronic and photonic dynamics are quantitatively captured by the Interacting-CWF approach, as shown in Fig. \ref{el_ph}, again with an extremely small trajectory ensemble.  

To summarize, we presented a method for solving the TDSE that is based on the recently introduced, exact, conditional decomposition of the many-body wavefunction\cite{PRL1}. We use the lowest order solution to the CWF equations of motion as a time-dependent basis, in a stochastic wavefunction ansatz which we call the Interacing-CWF approach. Our simulation results for the coupled electron-nuclear and electron-photon model system show that this method captures a quantitatively accurate physical picture, while using a number of trajectories that is orders of magnitude lower than the corresponding mean-field simulation. The degree of computational efficiency offered by this approach creates the possibility to treat dynamics in molecular and extended quantum systems with unprecedented accuracy without the need to pre-compute the BOPESs or NACTs, while providing access to all observables relevant for describing nonequilibrium dynamical phenomena.

In addition, these developments provide a general framework to approach the many-body problem in a variety of contexts. Notice that the decomposition of the full wave function offered here, in (\ref{CWF1}) and (\ref{CWF2}), is but one option of many possible conditional decompositions of the interacting many-body wavefunction. For example, using single-particle CWFs in a form compatible with time-dependent density functional theory is another particularly appealing route to follow in this respect, and work in this direction is already in progress. 
 
\section{Acknowledgements}
All of the authors thank Shunsuke Sato, Heiko Appel and Tarek A. Elsayed for fruitful discussions.
G.A. acknowledges financial support from the European  Union’s  Horizon 2020 research and innovation programme under the Marie Skłodowska-Curie grant agreement No 752822, the Spanish Ministerio de Economía y Competitividad (CTQ2016-76423-P), and the Generalitat de Catalunya (2017 SGR 348).
A.K. acknowledges funding from the National Sciences and Engineering Research Council of Canada Discovery grant program. 
A.R. acknowledges financial support from the European Research Council (ERC-2015-AdG-694097), and Grupos Consolidados (IT578-13). 

\bibliography{bibliography_PRL}

\begin{thebibliography}{44}
\expandafter\ifx\csname natexlab\endcsname\relax\def\natexlab#1{#1}\fi
\expandafter\ifx\csname bibnamefont\endcsname\relax
  \def\bibnamefont#1{#1}\fi
\expandafter\ifx\csname bibfnamefont\endcsname\relax
  \def\bibfnamefont#1{#1}\fi
\expandafter\ifx\csname citenamefont\endcsname\relax
  \def\citenamefont#1{#1}\fi
\expandafter\ifx\csname url\endcsname\relax
  \def\url#1{\texttt{#1}}\fi
\expandafter\ifx\csname urlprefix\endcsname\relax\def\urlprefix{URL }\fi
\providecommand{\bibinfo}[2]{#2}
\providecommand{\eprint}[2][]{\url{#2}}

\bibitem[{\citenamefont{Miller}(2012)}]{MillerPerspective}
\bibinfo{author}{\bibfnamefont{W.~H.} \bibnamefont{Miller}},
  \bibinfo{journal}{The Journal of chemical physics}
  \textbf{\bibinfo{volume}{136}}, \bibinfo{pages}{06B201}
  (\bibinfo{year}{2012}).

\bibitem[{\citenamefont{Tully}(2012)}]{TullyPerspective}
\bibinfo{author}{\bibfnamefont{J.~C.} \bibnamefont{Tully}},
  \bibinfo{journal}{J. Chem. Phys.} \textbf{\bibinfo{volume}{137}},
  \bibinfo{pages}{22A301} (\bibinfo{year}{2012}).

\bibitem[{\citenamefont{Kapral}(2015)}]{kapralreview}
\bibinfo{author}{\bibfnamefont{R.}~\bibnamefont{Kapral}},
  \bibinfo{journal}{Journal of Physics: Condensed Matter}
  \textbf{\bibinfo{volume}{27}}, \bibinfo{pages}{073201}
  (\bibinfo{year}{2015}).

\bibitem[{\citenamefont{McLachlan}(1964)}]{maclachlan}
\bibinfo{author}{\bibfnamefont{A.}~\bibnamefont{McLachlan}},
  \bibinfo{journal}{Molecular Physics} \textbf{\bibinfo{volume}{8}},
  \bibinfo{pages}{39} (\bibinfo{year}{1964}).

\bibitem[{\citenamefont{Tully}(1990)}]{tully90}
\bibinfo{author}{\bibfnamefont{J.~C.} \bibnamefont{Tully}},
  \bibinfo{journal}{The Journal of Chemical Physics}
  \textbf{\bibinfo{volume}{93}}, \bibinfo{pages}{1061} (\bibinfo{year}{1990}).

\bibitem[{\citenamefont{Martinez et~al.}(1996)\citenamefont{Martinez, Ben-Nun,
  and Levine}}]{FMS}
\bibinfo{author}{\bibfnamefont{T.~J.} \bibnamefont{Martinez}},
  \bibinfo{author}{\bibfnamefont{M.}~\bibnamefont{Ben-Nun}}, \bibnamefont{and}
  \bibinfo{author}{\bibfnamefont{R.}~\bibnamefont{Levine}},
  \bibinfo{journal}{The Journal of Physical Chemistry}
  \textbf{\bibinfo{volume}{100}}, \bibinfo{pages}{7884} (\bibinfo{year}{1996}).

\bibitem[{\citenamefont{Worth and Burghardt}(2003)}]{g-MCTDH}
\bibinfo{author}{\bibfnamefont{G.~A.} \bibnamefont{Worth}} \bibnamefont{and}
  \bibinfo{author}{\bibfnamefont{I.}~\bibnamefont{Burghardt}},
  \bibinfo{journal}{Chemical physics letters} \textbf{\bibinfo{volume}{368}},
  \bibinfo{pages}{502} (\bibinfo{year}{2003}).

\bibitem[{\citenamefont{Worth et~al.}(2004)\citenamefont{Worth, Robb, and
  Burghardt}}]{vmcg}
\bibinfo{author}{\bibfnamefont{G.}~\bibnamefont{Worth}},
  \bibinfo{author}{\bibfnamefont{M.}~\bibnamefont{Robb}}, \bibnamefont{and}
  \bibinfo{author}{\bibfnamefont{I.}~\bibnamefont{Burghardt}},
  \bibinfo{journal}{Faraday discussions} \textbf{\bibinfo{volume}{127}},
  \bibinfo{pages}{307} (\bibinfo{year}{2004}).

\bibitem[{\citenamefont{Shalashilin}(2010)}]{MCE}
\bibinfo{author}{\bibfnamefont{D.~V.} \bibnamefont{Shalashilin}},
  \bibinfo{journal}{The Journal of chemical physics}
  \textbf{\bibinfo{volume}{132}}, \bibinfo{pages}{244111}
  (\bibinfo{year}{2010}).

\bibitem[{\citenamefont{Makhov et~al.}(2014)\citenamefont{Makhov, Glover,
  Martinez, and Shalashilin}}]{cloning}
\bibinfo{author}{\bibfnamefont{D.~V.} \bibnamefont{Makhov}},
  \bibinfo{author}{\bibfnamefont{W.~J.} \bibnamefont{Glover}},
  \bibinfo{author}{\bibfnamefont{T.~J.} \bibnamefont{Martinez}},
  \bibnamefont{and} \bibinfo{author}{\bibfnamefont{D.~V.}
  \bibnamefont{Shalashilin}}, \bibinfo{journal}{The Journal of chemical
  physics} \textbf{\bibinfo{volume}{141}}, \bibinfo{pages}{054110}
  (\bibinfo{year}{2014}).

\bibitem[{\citenamefont{Sun and Miller}(1997)}]{scivr}
\bibinfo{author}{\bibfnamefont{X.}~\bibnamefont{Sun}} \bibnamefont{and}
  \bibinfo{author}{\bibfnamefont{W.~H.} \bibnamefont{Miller}},
  \bibinfo{journal}{The Journal of chemical physics}
  \textbf{\bibinfo{volume}{106}}, \bibinfo{pages}{6346} (\bibinfo{year}{1997}).

\bibitem[{\citenamefont{Miller}(2009)}]{fbivr}
\bibinfo{author}{\bibfnamefont{W.~H.} \bibnamefont{Miller}},
  \bibinfo{journal}{The Journal of Physical Chemistry A}
  \textbf{\bibinfo{volume}{113}}, \bibinfo{pages}{1405} (\bibinfo{year}{2009}).

\bibitem[{\citenamefont{Kapral and Ciccotti}(1999)}]{qcle}
\bibinfo{author}{\bibfnamefont{R.}~\bibnamefont{Kapral}} \bibnamefont{and}
  \bibinfo{author}{\bibfnamefont{G.}~\bibnamefont{Ciccotti}},
  \bibinfo{journal}{The Journal of chemical physics}
  \textbf{\bibinfo{volume}{110}}, \bibinfo{pages}{8919} (\bibinfo{year}{1999}).

\bibitem[{\citenamefont{Kelly et~al.}(2012)\citenamefont{Kelly, van Zon,
  Schofield, and Kapral}}]{mqcle}
\bibinfo{author}{\bibfnamefont{A.}~\bibnamefont{Kelly}},
  \bibinfo{author}{\bibfnamefont{R.}~\bibnamefont{van Zon}},
  \bibinfo{author}{\bibfnamefont{J.}~\bibnamefont{Schofield}},
  \bibnamefont{and} \bibinfo{author}{\bibfnamefont{R.}~\bibnamefont{Kapral}},
  \bibinfo{journal}{The Journal of Chemical Physics}
  \textbf{\bibinfo{volume}{136}}, \bibinfo{pages}{084101}
  (\bibinfo{year}{2012}).

\bibitem[{\citenamefont{Hsieh and Kapral}(2012)}]{jfbts}
\bibinfo{author}{\bibfnamefont{C.-Y.} \bibnamefont{Hsieh}} \bibnamefont{and}
  \bibinfo{author}{\bibfnamefont{R.}~\bibnamefont{Kapral}},
  \bibinfo{journal}{The Journal of chemical physics}
  \textbf{\bibinfo{volume}{137}}, \bibinfo{pages}{22A507}
  (\bibinfo{year}{2012}).

\bibitem[{\citenamefont{Shi and Geva}(2003)}]{ShiGeva-linear}
\bibinfo{author}{\bibfnamefont{Q.}~\bibnamefont{Shi}} \bibnamefont{and}
  \bibinfo{author}{\bibfnamefont{E.}~\bibnamefont{Geva}}, \bibinfo{journal}{The
  Journal of chemical physics} \textbf{\bibinfo{volume}{118}},
  \bibinfo{pages}{8173} (\bibinfo{year}{2003}).

\bibitem[{\citenamefont{Bonella and Coker}(2005)}]{landmap}
\bibinfo{author}{\bibfnamefont{S.}~\bibnamefont{Bonella}} \bibnamefont{and}
  \bibinfo{author}{\bibfnamefont{D.}~\bibnamefont{Coker}},
  \bibinfo{journal}{The Journal of chemical physics}
  \textbf{\bibinfo{volume}{122}}, \bibinfo{pages}{194102}
  (\bibinfo{year}{2005}).

\bibitem[{\citenamefont{Dunkel et~al.}(2008)\citenamefont{Dunkel, Bonella, and
  Coker}}]{ildm}
\bibinfo{author}{\bibfnamefont{E.}~\bibnamefont{Dunkel}},
  \bibinfo{author}{\bibfnamefont{S.}~\bibnamefont{Bonella}}, \bibnamefont{and}
  \bibinfo{author}{\bibfnamefont{D.}~\bibnamefont{Coker}},
  \bibinfo{journal}{The Journal of chemical physics}
  \textbf{\bibinfo{volume}{129}}, \bibinfo{pages}{114106}
  (\bibinfo{year}{2008}).

\bibitem[{\citenamefont{Huo and Coker}(2012)}]{pldm}
\bibinfo{author}{\bibfnamefont{P.}~\bibnamefont{Huo}} \bibnamefont{and}
  \bibinfo{author}{\bibfnamefont{D.~F.} \bibnamefont{Coker}},
  \bibinfo{journal}{The Journal of chemical physics}
  \textbf{\bibinfo{volume}{137}}, \bibinfo{pages}{22A535}
  (\bibinfo{year}{2012}).

\bibitem[{\citenamefont{Abedi et~al.}(2010)\citenamefont{Abedi, Maitra, and
  Gross}}]{exact_fac}
\bibinfo{author}{\bibfnamefont{A.}~\bibnamefont{Abedi}},
  \bibinfo{author}{\bibfnamefont{N.~T.} \bibnamefont{Maitra}},
  \bibnamefont{and} \bibinfo{author}{\bibfnamefont{E.~K.~U.}
  \bibnamefont{Gross}}, \bibinfo{journal}{Phys. Rev. Lett.}
  \textbf{\bibinfo{volume}{105}}, \bibinfo{pages}{123002}
  (\bibinfo{year}{2010}).

\bibitem[{\citenamefont{Min et~al.}(2015)\citenamefont{Min, Agostini, and
  Gross}}]{ct-mqc}
\bibinfo{author}{\bibfnamefont{S.~K.} \bibnamefont{Min}},
  \bibinfo{author}{\bibfnamefont{F.}~\bibnamefont{Agostini}}, \bibnamefont{and}
  \bibinfo{author}{\bibfnamefont{E.~K.} \bibnamefont{Gross}},
  \bibinfo{journal}{Physical review letters} \textbf{\bibinfo{volume}{115}},
  \bibinfo{pages}{073001} (\bibinfo{year}{2015}).

\bibitem[{\citenamefont{Ha et~al.}(2018)\citenamefont{Ha, Lee, and
  Min}}]{dish-xf}
\bibinfo{author}{\bibfnamefont{J.-K.} \bibnamefont{Ha}},
  \bibinfo{author}{\bibfnamefont{I.~S.} \bibnamefont{Lee}}, \bibnamefont{and}
  \bibinfo{author}{\bibfnamefont{S.~K.} \bibnamefont{Min}},
  \bibinfo{journal}{The journal of physical chemistry letters}
  \textbf{\bibinfo{volume}{9}}, \bibinfo{pages}{1097} (\bibinfo{year}{2018}).

\bibitem[{\citenamefont{Born and Oppenheimer}(1927)}]{BO}
\bibinfo{author}{\bibfnamefont{M.}~\bibnamefont{Born}} \bibnamefont{and}
  \bibinfo{author}{\bibfnamefont{R.}~\bibnamefont{Oppenheimer}},
  \bibinfo{journal}{Ann. Phys.} \textbf{\bibinfo{volume}{389}},
  \bibinfo{pages}{457} (\bibinfo{year}{1927}).

\bibitem[{\citenamefont{Domcke and Stock}(2007)}]{domcke}
\bibinfo{author}{\bibfnamefont{W.}~\bibnamefont{Domcke}} \bibnamefont{and}
  \bibinfo{author}{\bibfnamefont{G.}~\bibnamefont{Stock}},
  \emph{\bibinfo{title}{Theory of Ultrafast Nonadiabatic Excited-State
  Processes and their Spectroscopic Detection in Real Time}}
  (\bibinfo{publisher}{John Wiley \& Sons, Inc.}, \bibinfo{year}{2007}), pp.
  \bibinfo{pages}{1--169}.

\bibitem[{\citenamefont{Albareda et~al.}(2016)\citenamefont{Albareda, Abedi,
  Tavernelli, and Rubio}}]{PRA}
\bibinfo{author}{\bibfnamefont{G.}~\bibnamefont{Albareda}},
  \bibinfo{author}{\bibfnamefont{A.}~\bibnamefont{Abedi}},
  \bibinfo{author}{\bibfnamefont{I.}~\bibnamefont{Tavernelli}},
  \bibnamefont{and} \bibinfo{author}{\bibfnamefont{A.}~\bibnamefont{Rubio}},
  \bibinfo{journal}{Phys. Rev. A} \textbf{\bibinfo{volume}{94}},
  \bibinfo{pages}{062511} (\bibinfo{year}{2016}).

\bibitem[{\citenamefont{Albareda et~al.}(2014)\citenamefont{Albareda, Appel,
  Franco, Abedi, and Rubio}}]{PRL1}
\bibinfo{author}{\bibfnamefont{G.}~\bibnamefont{Albareda}},
  \bibinfo{author}{\bibfnamefont{H.}~\bibnamefont{Appel}},
  \bibinfo{author}{\bibfnamefont{I.}~\bibnamefont{Franco}},
  \bibinfo{author}{\bibfnamefont{A.}~\bibnamefont{Abedi}}, \bibnamefont{and}
  \bibinfo{author}{\bibfnamefont{A.}~\bibnamefont{Rubio}},
  \bibinfo{journal}{Phys. Rev. Lett.} \textbf{\bibinfo{volume}{113}},
  \bibinfo{pages}{083003} (\bibinfo{year}{2014}).

\bibitem[{\citenamefont{Bohm}(1952)}]{Bohm}
\bibinfo{author}{\bibfnamefont{D.}~\bibnamefont{Bohm}}, \bibinfo{journal}{Phys.
  Rev.} \textbf{\bibinfo{volume}{85}}, \bibinfo{pages}{166}
  (\bibinfo{year}{1952}).

\bibitem[{\citenamefont{Gindensperger et~al.}(2000)\citenamefont{Gindensperger,
  Meier, and Beswick}}]{gindensperger}
\bibinfo{author}{\bibfnamefont{E.}~\bibnamefont{Gindensperger}},
  \bibinfo{author}{\bibfnamefont{C.}~\bibnamefont{Meier}}, \bibnamefont{and}
  \bibinfo{author}{\bibfnamefont{J.}~\bibnamefont{Beswick}},
  \bibinfo{journal}{The Journal of Chemical Physics}
  \textbf{\bibinfo{volume}{113}}, \bibinfo{pages}{9369} (\bibinfo{year}{2000}).

\bibitem[{\citenamefont{Curchod et~al.}(2011)\citenamefont{Curchod, Tavernelli,
  and Rothlisberger}}]{Ivano}
\bibinfo{author}{\bibfnamefont{B.~F.~E.} \bibnamefont{Curchod}},
  \bibinfo{author}{\bibfnamefont{I.}~\bibnamefont{Tavernelli}},
  \bibnamefont{and}
  \bibinfo{author}{\bibfnamefont{U.}~\bibnamefont{Rothlisberger}},
  \bibinfo{journal}{Phys. Chem. Chem. Phys.} \textbf{\bibinfo{volume}{13}},
  \bibinfo{pages}{3231} (\bibinfo{year}{2011}).

\bibitem[{\citenamefont{Gu and Garashchuk}(2016)}]{Sophya}
\bibinfo{author}{\bibfnamefont{B.}~\bibnamefont{Gu}} \bibnamefont{and}
  \bibinfo{author}{\bibfnamefont{S.}~\bibnamefont{Garashchuk}},
  \bibinfo{journal}{The Journal of Physical Chemistry A}
  \textbf{\bibinfo{volume}{120}}, \bibinfo{pages}{3023} (\bibinfo{year}{2016}).

\bibitem[{\citenamefont{Albareda et~al.}(2015)\citenamefont{Albareda, Bofill,
  Tavernelli, Huarte-Larrañaga, Illas, and Rubio}}]{JPCL1}
\bibinfo{author}{\bibfnamefont{G.}~\bibnamefont{Albareda}},
  \bibinfo{author}{\bibfnamefont{J.~M.} \bibnamefont{Bofill}},
  \bibinfo{author}{\bibfnamefont{I.}~\bibnamefont{Tavernelli}},
  \bibinfo{author}{\bibfnamefont{F.}~\bibnamefont{Huarte-Larrañaga}},
  \bibinfo{author}{\bibfnamefont{F.}~\bibnamefont{Illas}}, \bibnamefont{and}
  \bibinfo{author}{\bibfnamefont{A.}~\bibnamefont{Rubio}},
  \bibinfo{journal}{The Journal of Physical Chemistry Letters}
  \textbf{\bibinfo{volume}{6}}, \bibinfo{pages}{1529} (\bibinfo{year}{2015}).

\bibitem[{\citenamefont{{Elsayed} et~al.}(2017)\citenamefont{{Elsayed},
  {M{\o}lmer}, and {Bojer Madsen}}}]{Tarek}
\bibinfo{author}{\bibfnamefont{T.~A.} \bibnamefont{{Elsayed}}},
  \bibinfo{author}{\bibfnamefont{K.}~\bibnamefont{{M{\o}lmer}}},
  \bibnamefont{and} \bibinfo{author}{\bibfnamefont{L.}~\bibnamefont{{Bojer
  Madsen}}}, \bibinfo{journal}{ArXiv e-prints}  (\bibinfo{year}{2017}),
  \eprint{1706.00818}.

\bibitem[{\citenamefont{Benseny et~al.}(2014)\citenamefont{Benseny, Albareda,
  Sanz, Mompart, and Oriols}}]{EPJD}
\bibinfo{author}{\bibfnamefont{A.}~\bibnamefont{Benseny}},
  \bibinfo{author}{\bibfnamefont{G.}~\bibnamefont{Albareda}},
  \bibinfo{author}{\bibfnamefont{{\'A}.~S.} \bibnamefont{Sanz}},
  \bibinfo{author}{\bibfnamefont{J.}~\bibnamefont{Mompart}}, \bibnamefont{and}
  \bibinfo{author}{\bibfnamefont{X.}~\bibnamefont{Oriols}},
  \bibinfo{journal}{The European Physical Journal D}
  \textbf{\bibinfo{volume}{68}}, \bibinfo{pages}{286} (\bibinfo{year}{2014}).

\bibitem[{\citenamefont{Oriols}(2007)}]{PRLoriols}
\bibinfo{author}{\bibfnamefont{X.}~\bibnamefont{Oriols}},
  \bibinfo{journal}{Phys. Rev. Lett.} \textbf{\bibinfo{volume}{98}},
  \bibinfo{pages}{066803} (\bibinfo{year}{2007}).

\bibitem[{\citenamefont{Beck et~al.}(2000)\citenamefont{Beck, Jäckle, Worth,
  and Meyer}}]{MCTDH_review}
\bibinfo{author}{\bibfnamefont{M.}~\bibnamefont{Beck}},
  \bibinfo{author}{\bibfnamefont{A.}~\bibnamefont{Jäckle}},
  \bibinfo{author}{\bibfnamefont{G.}~\bibnamefont{Worth}}, \bibnamefont{and}
  \bibinfo{author}{\bibfnamefont{H.-D.} \bibnamefont{Meyer}},
  \bibinfo{journal}{Phys. Rep.} \textbf{\bibinfo{volume}{324}},
  \bibinfo{pages}{1 } (\bibinfo{year}{2000}).

\bibitem[{({\natexlab{a}})}]{SIA}
\bibinfo{note}{See Appendix A in the SI for a detailed description of the
  numerical implementation of the ICWF method including specific expressions to
  compute the electronic BO populations, the reduced nuclear coordinate
  density, and the decoherence indicator in terms of the ansatz in
  \eqref{ansatz}.}

\bibitem[{\citenamefont{Shin and Metiu}(1995)}]{Metiu}
\bibinfo{author}{\bibfnamefont{S.}~\bibnamefont{Shin}} \bibnamefont{and}
  \bibinfo{author}{\bibfnamefont{H.}~\bibnamefont{Metiu}}, \bibinfo{journal}{J.
  Chem. Phys.} \textbf{\bibinfo{volume}{102}}, \bibinfo{pages}{9285}
  (\bibinfo{year}{1995}).

\bibitem[{({\natexlab{b}})}]{SIB}
\bibinfo{note}{See Appendix B in the SI for a detailed description of the
  Shin-Metiu model and the specific parameters used in this work.}

\bibitem[{\citenamefont{Abedi et~al.}(2013)\citenamefont{Abedi, Agostini,
  Suzuki, and Gross}}]{Ali2}
\bibinfo{author}{\bibfnamefont{A.}~\bibnamefont{Abedi}},
  \bibinfo{author}{\bibfnamefont{F.}~\bibnamefont{Agostini}},
  \bibinfo{author}{\bibfnamefont{Y.}~\bibnamefont{Suzuki}}, \bibnamefont{and}
  \bibinfo{author}{\bibfnamefont{E.~K.~U.} \bibnamefont{Gross}},
  \bibinfo{journal}{Phys. Rev. Lett.} \textbf{\bibinfo{volume}{110}},
  \bibinfo{pages}{263001} (\bibinfo{year}{2013}).

\bibitem[{\citenamefont{Agostini et~al.}(2015)\citenamefont{Agostini, Abedi,
  Suzuki, Min, Maitra, and Gross}}]{fede}
\bibinfo{author}{\bibfnamefont{F.}~\bibnamefont{Agostini}},
  \bibinfo{author}{\bibfnamefont{A.}~\bibnamefont{Abedi}},
  \bibinfo{author}{\bibfnamefont{Y.}~\bibnamefont{Suzuki}},
  \bibinfo{author}{\bibfnamefont{S.~K.} \bibnamefont{Min}},
  \bibinfo{author}{\bibfnamefont{N.~T.} \bibnamefont{Maitra}},
  \bibnamefont{and} \bibinfo{author}{\bibfnamefont{E.~K.~U.}
  \bibnamefont{Gross}}, \bibinfo{journal}{The Journal of Chemical Physics}
  \textbf{\bibinfo{volume}{142}}, \bibinfo{pages}{084303}
  (\bibinfo{year}{2015}).

\bibitem[{\citenamefont{Flick et~al.}(2015)\citenamefont{Flick, Ruggenthaler,
  Appel, and Rubio}}]{PNAS1}
\bibinfo{author}{\bibfnamefont{J.}~\bibnamefont{Flick}},
  \bibinfo{author}{\bibfnamefont{M.}~\bibnamefont{Ruggenthaler}},
  \bibinfo{author}{\bibfnamefont{H.}~\bibnamefont{Appel}}, \bibnamefont{and}
  \bibinfo{author}{\bibfnamefont{A.}~\bibnamefont{Rubio}},
  \bibinfo{journal}{Proceedings of the National Academy of Sciences}
  \textbf{\bibinfo{volume}{112}}, \bibinfo{pages}{15285}
  (\bibinfo{year}{2015}).

\bibitem[{\citenamefont{Flick et~al.}(2017)\citenamefont{Flick, Ruggenthaler,
  Appel, and Rubio}}]{PNAS2}
\bibinfo{author}{\bibfnamefont{J.}~\bibnamefont{Flick}},
  \bibinfo{author}{\bibfnamefont{M.}~\bibnamefont{Ruggenthaler}},
  \bibinfo{author}{\bibfnamefont{H.}~\bibnamefont{Appel}}, \bibnamefont{and}
  \bibinfo{author}{\bibfnamefont{A.}~\bibnamefont{Rubio}},
  \bibinfo{journal}{Proceedings of the National Academy of Sciences}
  \textbf{\bibinfo{volume}{114}}, \bibinfo{pages}{3026} (\bibinfo{year}{2017}).

\bibitem[{\citenamefont{Ruggenthaler et~al.}(2018)\citenamefont{Ruggenthaler,
  Tancogne-Dejean, Flick, Appel, and Rubio}}]{Nature}
\bibinfo{author}{\bibfnamefont{M.}~\bibnamefont{Ruggenthaler}},
  \bibinfo{author}{\bibfnamefont{N.}~\bibnamefont{Tancogne-Dejean}},
  \bibinfo{author}{\bibfnamefont{J.}~\bibnamefont{Flick}},
  \bibinfo{author}{\bibfnamefont{H.}~\bibnamefont{Appel}}, \bibnamefont{and}
  \bibinfo{author}{\bibfnamefont{A.}~\bibnamefont{Rubio}},
  \bibinfo{journal}{Nature Reviews Chemistry} \textbf{\bibinfo{volume}{2}},
  \bibinfo{pages}{0118} (\bibinfo{year}{2018}).

\bibitem[{({\natexlab{c}})}]{SIC}
\bibinfo{note}{See Appendix C in the SI for a detailed description of the
  cavity bound electron-photon model system.}

\end{thebibliography}

\end{document}